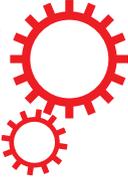

# SCIENTIFIC REPORTS

# Interface instability modes in freezing colloidal suspensions: revealed from onset of planar instability

Lilin Wang[1], Jiaxue You[2], Zhijun Wang[2], Jincheng Wang[2] & Xin Lin[2]

Freezing colloidal suspensions widely exists in nature and industry. Interface instability has attracted much attention for the understandings of the pattern formation in freezing colloidal suspensions. However, the interface instability modes, the origin of the ice banding or ice lamellae, are still unclear. *In-situ* experimental observation of the onset of interface instability remains absent up to now. Here, by directly imaging the initial transient stage of planar interface instability in directional freezing colloidal suspensions, we proposed three interface instability modes, Mullins-Sekerka instability, global split instability and local split instability. The intrinsic mechanism of the instability modes comes from the competition of the solute boundary layer and the particle boundary layer, which only can be revealed from the initial transient stage of planar instability in directional freezing.

It is very important and urgent to figure out the interface instability mechanisms of freezing colloidal suspensions, as the freezing of colloidal suspensions attracts more and more attention in the interdisciplinary researches of porous ceramics[1–3], solidification[4–6], geocryology[7,8], etc. For example, directional freezing of aqueous suspensions, also called ice-templating method, has been used to produce a variety of aligned porous structure materials with widespread applications, such as filtration, biomedical implant, catalytic carrier, fuel cell and micro-fluid[9,10]. The pore formation in the ice-templating method is determined by the ice growth during the freezing process, while the morphology and size of ice crystals greatly depend on the solid/liquid interface instability. As to the interface instability, a consensus of constitutional undercooling from Mullins-Sekerka (MS) instability has been addressed[11]. However, with a condensed particle layer in front of the interface of the freezing colloidal suspensions, the interface instability mechanism is encountering challenges. Ten years ago, by virtue of some fundamental knowledge of particle constitutional supercooling of the accumulated particles, morphological stability analysis of a planar interface, developed from solidification of alloy systems, has also been employed to understand the directional freezing of colloidal suspensions[12,13]. However, we demonstrated that the interface undercoolings in the freezing of colloidal suspensions mainly come from the solute constitutional supercooling rather than the particle constitutional supercooling very recently[14]. It raises the crisis that what contribution of accumulated particles is on the interface instability without the particle-induced constitutional supercooling.

The emerging research frontier of freezing colloidal suspensions also provides a new challenge to the topic of interface instability. As the beginning of the pattern formation, interface instability is a common phenomenon in various natural and industrial processes of self-organization patterning[15]. In the solidification, the interface instability has been well analyzed based on the linear stability analysis of MS instability. However, in the colloidal suspensions system, plenty of nano-particles are accumulated in front of the solid/liquid interface as the solvent of the suspensions transforms from liquid to solid. It is still not clear when and how the interface instability occurs in such phase transformation with complex interactions between the particles and the solid/liquid interface. Over the past decade, there have been arguments and conjectures on the solute effects and particle effects on the pattern formation of freezing of colloidal suspensions[12,16]. However, in spite of these efforts, the origin of the interface instability, one of the most important issues, was ignored. Investigation on the initial interface instability[17–20] will provide abundant information, and hence solve the puzzles of interface instability of colloidal suspensions.

[1]School of Materials Science and Engineering, Xi'an University of Technology, Xi'an 710048, P. R. China. [2]State Key Laboratory of Solidification Processing, Northwestern Polytechnical University, Xi'an 710072, P. R. China. Correspondence and requests for materials should be addressed to Z.W. (email: zhjwang@nwpu.edu.cn) or X.L. (email: xlin@nwpu.edu.cn)

It also should be noted that the ice banding phenomenon[21,22] has been seldom mentioned in the previous investigations on interface instability during freezing colloidal suspensions. The ice banding is a common phenomenon in the freezing of soil, and has been reproduced in laboratory. However, the forming mechanisms of ice banding have been investigated without considering the interface instability. The relationships between the ice banding and interface instability are also need to be clarified.

In this letter, we revealed the secrets of interface instability in the freezing colloidal suspensions by focusing on the onset of interface instability through a well-designed directional freezing experimental apparatus[23]. Different interface instability morphologies were *in-situ* observed. Three interface instability modes are proposed based on the analyses of establishing boundary layers of solutes and particles ahead of the interface. The intrinsic mechanism of instability modes is also proved by well-designed experiments.

The experiments were carried out in a high precision directional solidification apparatus, which has been used to quantitatively measure the interface undercooling in the freezing of colloidal suspensions[23]. The innovation of the apparatus is the *in-situ* comparison of the solid/liquid interfaces of colloidal suspensions and its supernatant. Here, we compared the dynamic evolution of the interfaces of the colloidal suspensions and its supernatant during the planar interface instability of directional solidification. Colloidal suspensions of $\alpha-$alumina powder with mean diameter d = 50 nm (Wanjing New Material, Hangzhou, China, ≥99.95% purity, monodispersity) were prepared by using HCl (hydrogen chloride) and deionized water. Also the stable dispersion of alumina suspensions has been confirmed[21]. The particles had a density of $3.97\,\mathrm{g\,cm^{-3}}$. Three systems with different initial volume fractions of particles, $\phi_0 = 1.31\%$, 3.63%, 7.75% (wt = 5%, 13%, 25%), were designed to reveal the particle accumulation effects for different volume fractions. Before pulling, the system is homogenized for one hour. The interface morphologies and positions were recorded by one frame per second. The pulling speed of V = 16 μm/s and the thermal gradient of G = 7.23 K/cm is constant only in the experiments of revealing the instability modes.

## Onset of Interface Instability and Instability Modes

With pulling velocity larger than a critical instability velocity, the planar interface undergoes instability process[20]. The critical interface instability morphologies in different systems are focused to reveal the interface instability modes in the freezing of colloidal suspensions. Figure 1 shows three typical onset morphologies of the instability of planar interface in different systems. The adjacent cells are the colloidal suspensions and its supernatant, respectively. The black dot lines represent the initial solid/liquid interfaces after homogenization, moving with the samples. The entire processes from the very beginning of pulling to onset of interface instability are shown in the supplemental videos (Movies S1–S3) for different systems. Movies S1, S2 and S3 correspond to $\phi_0 = 1.31\%$, 3.63% and 7.75%, respectively.

For the supernatant, fluctuation of small amplitude appears on the planar interface after an incubation time, as shown in Fig. 1. The amplitude enlarged rapidly to a finite level to form cellular structure after the instability, shown as Fig. S1 (Supplemental Material). The interface instability of the supernatant obeys the classical MS instability dynamics which has been well predicted by time-dependent instability analysis[17,20]. However, the instability processes of the colloidal suspensions are of great difference from the supernatant system and depend greatly on their particle volume fractions.

The directional freezing of the colloidal suspensions with small particle volume fraction undergoes the similar process as that of the supernatant, as shown in Fig. 1(a). The interface instability also starts from the fluctuation and then develops into cellular structure. The visible fluctuation occurs almost at the same time of that in the supernatant cell, only a little earlier. In this system, the accumulated particle layer has a little bit impact on the incubation time of planar instability. However, as the volume fraction increases, the instability mode totally changes. As shown in Fig. 1(b), the cellular instability disappears. Instead, the accumulated particle layer is split and trapped in the ice at the onset of planar instability of colloidal suspensions. Moreover, this kind of instability happens much earlier than the cellular instability mode. The split comes from the local insertion of ice spears, indicated by the bright spots and the protrusion marked in Fig. 1(b). We called this as "local split instability". The definition is more easily to be understood based on the morphology of steady growth, as expounded in the supplemental material.

As the volume fraction further increases, the accumulated particle layer is split into stripe bands at the beginning of the planar instability as shown in Fig. 1(c). The instability mode is similar to the local split instability mode shown in Fig. 1(b), but here the split block is a stripe band jointed with ice lens. The spears penetrate the accumulated particle layer and then grow laterally to form the ice lens. We called it as "global split instability". This kind of instability mode is almost irrelevant from the MS instability of its supernatant. The accumulated particle boundary layer will split even that the planar interface of the supernatant is stable with a pulling velocity smaller than the critical one.

## The Origin of the Interface Instability

The interface instability modes exhibit distinct characteristics, indicating different instability mechanisms. The intrinsic factors, determining the interface instability, need to be further found out. As shown in the MS instability analysis, the interface instability is related to the solute boundary layer ahead of the planar interface[17]. However, there are two types of boundary layers ahead of the planar interface in the freezing of colloidal suspensions, i.e., solute boundary layer and particle boundary layer. The linear stability analysis was also used to analyze the morphological stability with a particle boundary layer[24,25]. Therefore, the establishing processes of these two kinds of boundary layers in the initial transient stage are very interesting and helpful to illustrate the interface instability behaviors.

It is very difficult to directly observe the establishing of solute boundary layer and particle boundary layer. However, the dynamic establishing of the two boundary layers can be schematically presented based on the time-dependent analysis of interface migration and the preliminary precise experimental results of solute boundary layer establishing in directional solidification[26]. For one dimensional free boundary diffusion problem, the profiles of solute concentration and particle volume fraction along the system are schematically shown in Fig. 2(a,b) respectively. Figure 2(a) is a well-known process of solidification[27]. As the planar interface propagates, the ejected solutes and particles accumulate in front of the interface and diffuse into the liquid away from the interface. Accordingly, the solute and particle profiles across the system can be solved by time-dependent diffusion equations and boundary conditions at the solid/liquid interface. The time-dependent solute concentration and particle volume fraction in front of the solid/liquid interface is shown by connecting the interface concentration at the liquid side, as shown by the black thick lines in Fig. 2(a,b). The solute concentration in front of the interface will increase from $C_0$ to $C_0/k$ gradually in the initial transient stage. $C_0$ is the initial solute concentration in the solvent of suspensions and $k$ is the partition coefficient of solutes. While for the particle, since the diffusion constant

$$D_p = k_B T / 6\pi r \eta \qquad (1)$$

is around $10^{-11}$ m²/s for particle with radius of 50 nm at ambient temperature, where $k_B$ is Boltzmann constant, $r$ is the radius of the particles, $\eta$ is the viscosity and $T$ is temperature. Compared with the solute diffusion constant, the particle diffusion constant is much smaller of two orders of magnitude. The particle concentration in front of interface will rapidly increase to the maximum volume fraction of particles $\phi_{max}$ within a very short time $t_s$, as shown in Fig. 2(b). Furthermore, the length scale of the equivalent particle diffused layer can be ignored, compared with that of the solute diffused layer, indicating that diffusion of particles is negligible[22].

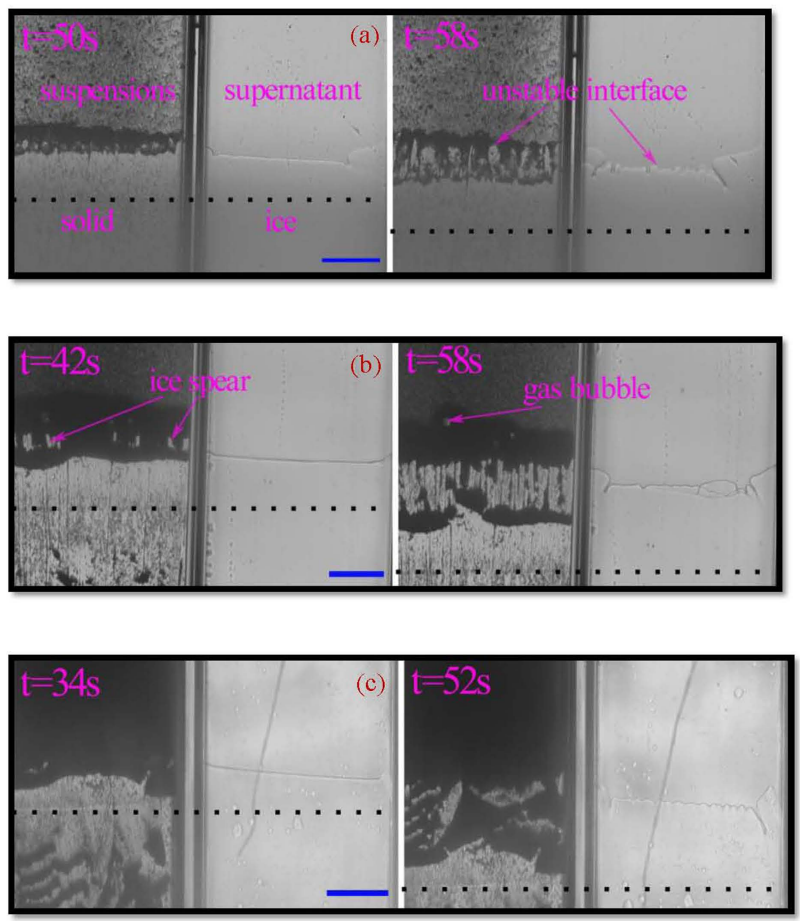

**Figure 1.** Onset of the planar interface instability in colloidal suspensions with different volume fractions of particles, (**a**) $\phi_0 = 1.31\%$, (**b**) $\phi_0 = 3.63\%$, (**c**) $\phi_0 = 7.75\%$. The control parameters are V = 16 μm/s, G = 7.23 K/cm in the systems. The scale bar is 300 μm. For each frame, there are two adjacent cells contain colloidal suspensions and its supernatant. The early one shows the onset of the interface instability of the colloidal suspensions. The later one shows the onset of the interface instability of the supernatant. The ice spear in the split instability mode are marked in (**b**).

The boundary layers of the solute and the particle ahead of the interface are totally different. The solute boundary layer profile decays exponentially in the colloidal suspensions, while the particle boundary layer profile shows platform of $\phi_{max}$ and then rapidly decreases to the initial volume fraction $\phi_0$, as shown in Fig. 2. The discrepancies of these two profiles mainly come from the differences of diffusion constants. The establishing boundary layer of solute has been well described in the time-dependent MS instability analysis[17], with a the diffusion length of $l_d = 2D_s/V_I$, where $D_s$ is the diffusivity of the solute and $V_I$ the instantaneous interface velocity. The diffusion length is one of the main factors determining the MS instability. For the particle boundary layer, the width of the accumulated particle layer $W$ should also be responsible for the interface instability,

$$W = \frac{\int \phi_0 V_I dt}{\phi_{max}}. \qquad (2)$$

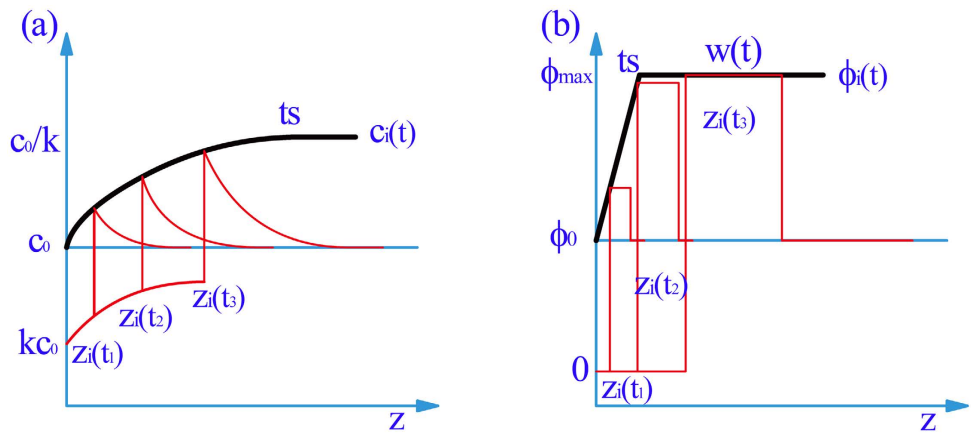

**Figure 2.** Sketches of the time-dependent solute concentration (**a**) and particle volume fraction (**b**) in front of the solid/liquid interface (black thick lines). The red thin lines are the profiles of solute (**a**) and particle boundaries (**b**) for a given time. $t_s$ is the time reaching steady state. $C_0$ is the initial solute concentration, $C_i$ is the liquid solute concentration in the front of solid/liquid interface, $z_i$ is the interface position, k is the partition coefficient. $\phi_0$ is the initial particle volume fraction, $\phi_i$ is the particle volume fraction in the front of solid/liquid interface, $\phi_{max}$ is the maximum particle volume fraction in the front of solid/liquid interface. w(t) is the width of the particle layer.

The width of the particle layer depends on the instantaneous interface velocity and the initial volume fraction of the particles. The solute boundary layer will induce MS instability if there is a constitutional undercooling region in front of the solid/liquid interface. For the particle boundary layer, the constitutional undercooling of thermodynamics is much smaller compared with the positive thermal gradient. However, the steadily increasing of particle layer width ahead of the planar interface will prevent the migration of the planar interface by holding back the interstitial flow of water, and the new ice lens may forms[28], indicating the instability of interface. This case corresponds to the split instability.

Considering the accumulated boundary layers of solute and particle, the three interface instability modes observed in Fig. 1 can be well understood. 1) The cellular instability is determined by the MS instability from the accumulated solute boundary layer. After interface instability, particles are submerged into intercellular space and then the particle layer is within a limited width. The onset of the initial interface instability of colloidal suspensions synchronizes with the interface instability of supernatant, where the solute effect is dominant compared with the effect of particles. 2) The global split instability mode is mainly determined by the accumulated particle boundary layer of large volume fraction system, related to the forming mechanism of ice lens. The whole interface is entrapped before the MS instability of solute boundary layer. In the global split instability, the particle boundary layer is dominant. 3) When the effects of the particle boundary layer and the solute boundary layer are at the same level, the local split instability mode occurs.

The variation of interface position in the transient stage is the key issue to reflect the establishing of the diffusion boundary layer[17,20], and can be used to reveal the interface instability modes in the freezing colloidal suspensions. The interface position evolutions with time during the transient stage are shown in Fig. 3. The interface instability moments of the supernatant and colloidal suspensions are marked in the curves. Before interface instability, the movements of interfaces in the two adjacent cells are the same for all the three systems with different volume fractions of particles, as shown in Fig. 3. Although there is a dense particle layer in front of the interface, the migration of interface position is almost independent of the particle layer compared with the interface position migration of the supernatant. It indicates that the dense particle layer with a finite width in front of the interface has little impact on the interface evolution. Instead, the accumulated solute boundary layer in both of the supernatant and colloidal suspensions determines the interface migration behavior before interface instability. The interface undercooling from solute constitutional redistribution mainly determines the interface migration before interface instability. The comparison of the interface positions in the systems of the colloidal suspensions and its supernatant also proves the absence of particle-induced interface undercooling[14]. After interface instability, the interface evolves differently in different instability modes. For the MS instability and the local split modes, the undercooling of cellular tip is constant. However, for the global split mode, the solid/liquid interface undercooling oscillates.

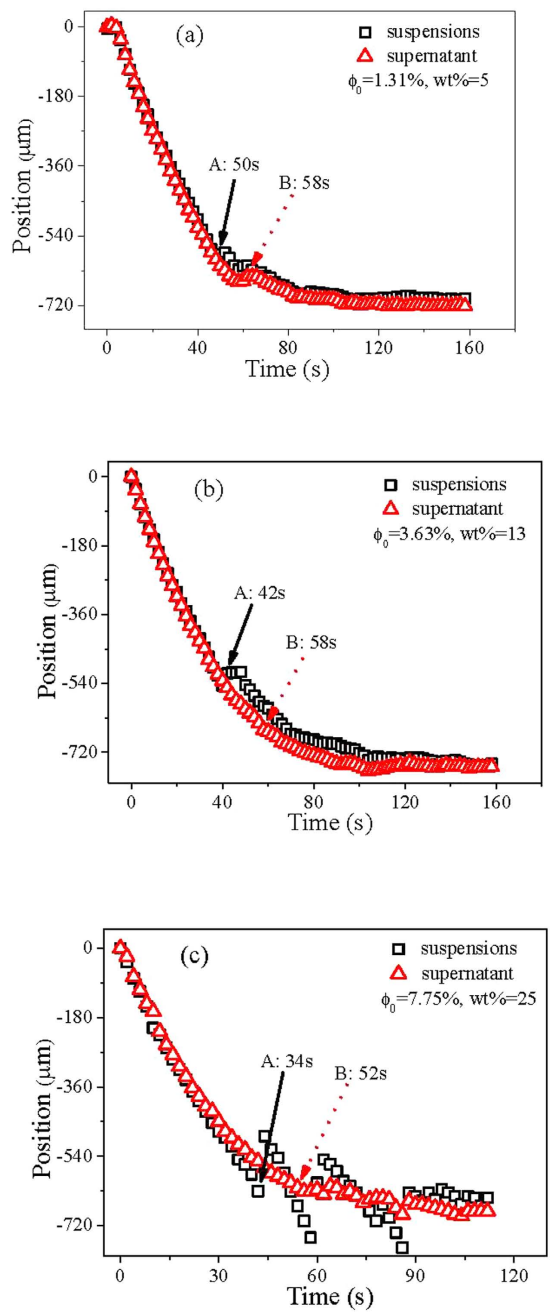

**Figure 3. The interface migration in the three systems with different volume fractions of particles.** The interface instability times (A for colloidal suspensions, B for supernatant) are marked. Before interface instability, the interface migrations of the colloidal suspension and supernatant synchronize. The interface oscillation in (**c**) corresponds to the periodical formation of ice lens.

### The Parameters Determining the Interface Instability Modes

The selection of the instability modes depends great on the competition of the effects of solute in the supernatant and the condensed particle layer in front of the interface. More details of the intrinsic mechanism of the instability modes will be presented in this section based on experimental evidences: the absence of particle effects on the cellular instability, the pulling velocity dependent interface instability and the variation of interface instability with additive.

Although the particle effects on the interface instability are controversial in freezing colloidal suspensions, the absence of particle effects on cellular instability can be clarified. In the pattern formation of directional solidification, it is accepted that the MS instability causes the cellular morphology. The linear stability analysis of particle induced constitutional undercooling is a replication of the MS instability. However, the particle induced constitutional undercooling is smaller than $10^{-5}$ K and there is almost no particle diffusion boundary layer in front of the solid/liquid planar interface. Moreover, the previous experimental results[29] indicated that the absence of cellular structure in the cases of minor additives, but the cellular structure appears with the increasing of additives. Therefore, the particles in the suspensions alone cannot induce the cellular instability, but cause ice lenses.

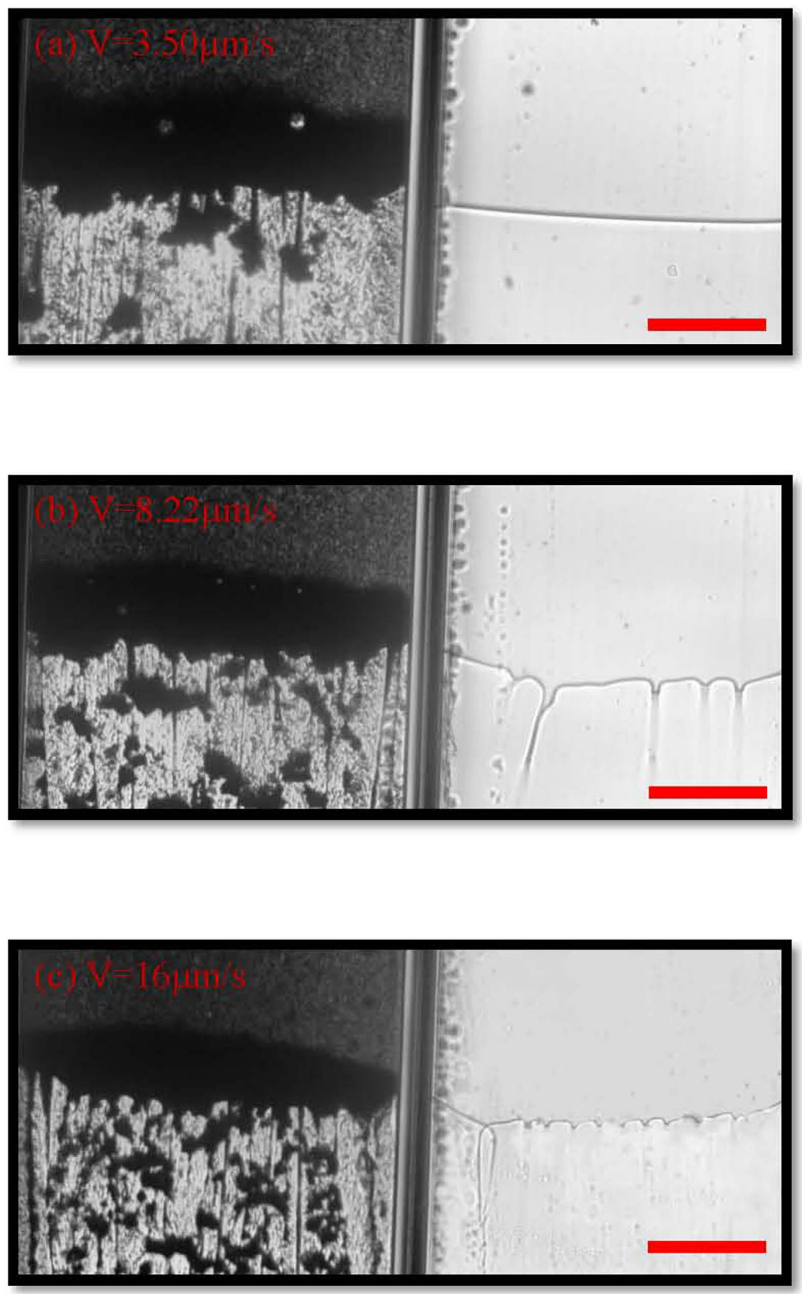

**Figure 4.** Steady-state growth of suspensions and the supernatant with different pulling speeds (**a**) V = 3.50 μm/s, (**b**) V = 8.22 μm/s and (**c**) V = 16 μm/s. The control parameters are volume fractions of particles $\phi_0$ = 3.63% and G = 7.23 K/cm in the systems. The scale bar is 300 μm. For each frame, there are two adjacent cells contain colloidal suspensions (left) and its supernatant (right).

The MS instability only happens at a velocity larger than the critical criterion in directional solidification[27]. Based on this, we design experiments with different pulling velocities. The steady interface morphologies of different pulling velocities are shown in Fig. 4. At a smaller velocity of 3.5 μm/s, the planar interface is stable in the supernatant system, while the local split instability occurs in the colloidal suspensions system, as shown in Fig. 4(a). As the pulling velocity larger than the critical one, the interface propagates with cellular morphology in the supernatant system, and the local split particle clusters becomes smaller and smaller in the colloidal suspensions system, as shown in Fig. 4(b,c). The experimental results indicate that the increasing solute effects gradually change the local split morphologies and may induces into cellular structures as the pulling velocity further increases.

As the solute effects can greatly enhances the planar instability, increasing the content of additives will change the interface morphologies. We found recent experiments regarding to the effects of additives[29]. Delattre *et al.*

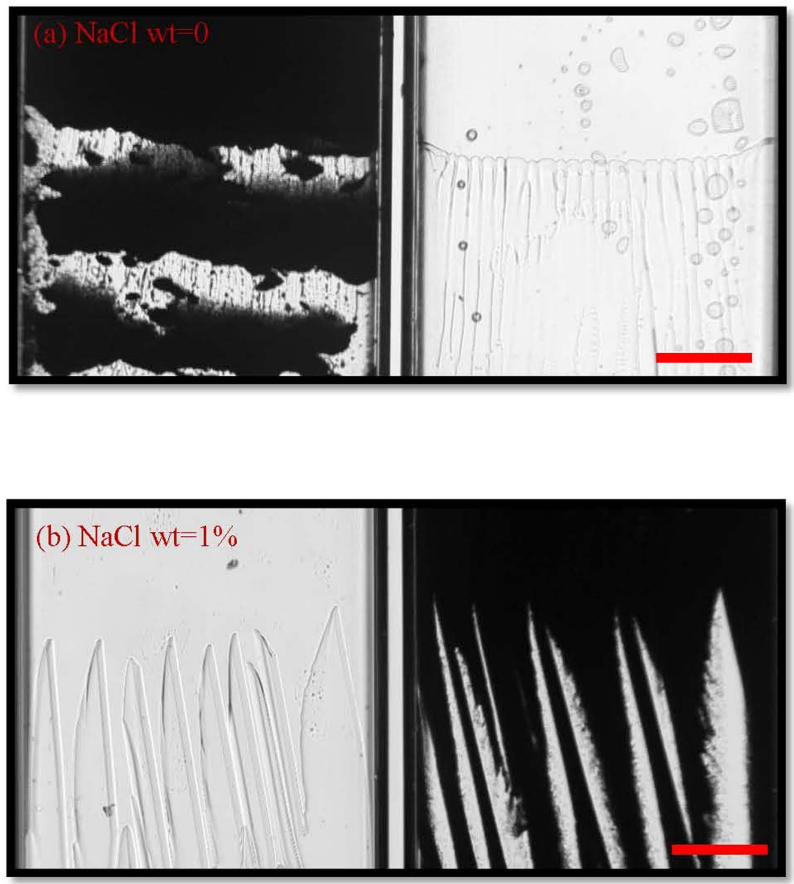

**Figure 5.** Steady-state growth of suspensions and the supernatant with different mass content of sodium chloride (**a**) wt = 0 and (**b**) wt = 1% of particle amount. The control parameters are volume fractions of particles $\phi_0 = 9.74\%$, V = 16 μm/s and G = 7.23 K/cm in the systems. The scale bar is 300 μm. For each frame, there are two adjacent cells contain colloidal suspensions and its supernatant.

confirmed the transition from local split mode to the cellular instability mode by increasing the content of additives by *in situ* X-ray images. They attributed the transition to the variation of viscosity and other factors related to the absorption effects of particle interface. However, the transition of interface morphologies may come from the increasing of solute effects on the interface instability. According to the understanding of competition of solute effects and particle effects, we added NaCl into our system. The addition of NaCl almost does not change the viscosity of the system. Moreover, the NaCl is an inorganic small molecular compound, neither a binder nor dispersant. The experimental results are shown in Fig. 5. It shows that the global split instability mode in Fig. 5(a) is totally changed into the cellular instability mode by only a small amount of NaCl (1%wt of particle amount). Accordingly, it is convinced that the transition from ice banding to cellular structure comes from the increasing of solute effects on the interface instability.

Although we have proved some important factors determining the interface instability modes based on preliminary experiment results, there are many more factors affecting the MS instability and the global split instability, including intrinsic physical parameters and control parameters of freezing. In general, as to the MS instability, there are solute diffusion constants, solute partition coefficient, slope of liquidus line, surface tension, initial solute concentration, thermal gradient, pulling velocity etc[11]. For the global split instability, there are particle diffusion constants, particle partition coefficient, particle size distribution, particle shape, initial particle volume fraction, viscosity, thermal gradient, pulling velocity, etc. The interface instability mode can be controlled by adjusting these control factors or selecting materials. Although it is very complex, some clues can be used to guide the design process. For example, the MS instability will win by reducing the initial particle concentration or increasing the solute concentration. On the other hand, the global split mode can win by reducing the solute concentration in the solvent of the suspensions or increasing the particle volume fraction.

## Conclusions

The planar interface instability in the directional solidification of colloidal suspensions has been investigated through the *in situ* observation of transient stage of the initial planar instability for the first time. The novel *in situ* comparison of supernatant system and colloidal suspensions reveals the mechanism of planar instability of freezing colloidal suspensions. Three different instability modes were reported, the cellular MS instability, local split mode and global split mode. The cellular structure instability forms the cellular or lamellae structure. The local split mode presents the cellular structure plus trapped clusters. The global split mode forms the band structure.

During the instability process, the interface position evolution is determined by the solute redistribution ahead of the interface, and the planar interface losses its stability earlier in the local and global split modes than that in the cellular instability mode. The designed experiments proved that the selection of the instability modes depends greatly on the competition of the effects of solute in the supernatant and the condensed particle layer in front of the interface.

## Acknowledgements

This research has been supported by Nature Science Foundation of China (Grant No. 51371151), Free Research Fund of State Key Laboratory of Solidification Processing (100-QP-2014), the Fund of State Key Laboratory of Solidification Processing in NWPU (13-BZ-2014) and the Fundamental Research Funds for the Central Universities (3102015ZY020).

## Author Contributions

L.W. and Z.W. designed the research, J.Y. performed the experiments, J.W. and X.L. processed experimental data, all authors analyzed and discussed the results, L.W. and Z.W. wrote the paper.

## Additional Information

**Supplementary information** accompanies this paper at http://www.nature.com/srep

**Competing financial interests:** The authors declare no competing financial interests.

**How to cite this article**: Wang, L. *et al.* Interface instability modes in freezing colloidal suspensions: revealed from onset of planar instability. *Sci. Rep.* **6,** 23358; doi: 10.1038/srep23358 (2016).